\documentclass[prb,showpacs,twocolumn]{revtex4}
\usepackage{amsmath,amsfonts,amsthm}
\usepackage{pstricks,pst-coil}
\newcommand{\eq}[1]{(\ref{#1})}  

\begin{document}
\title{A model with simultaneous\\ first and second order phase transitions}
\author{Alain Messager}\email{messager@cpt.univ-mrs.fr}
\affiliation{Centre de Physique Th\'eorique, CNRS--Luminy,
Case 907, F-13288 Marseille, France}
\author{Bruno Nachtergaele}\email{bxn@math.ucdavis.edu}
\affiliation{Department of Mathematics, University of California at Davis, 
Davis, CA 95616}
\date{\today}
\begin{abstract}
We introduce a two dimensional nonlinear XY model with
a {\it second order phase transition driven by spin waves}, together with 
a {\it first order phase transition} in the bond variables between two ``bond ordered phases'', 
one with local ferromagnetic order and another with local anti-ferromagnetic order.
We also prove that at the transition temperature the bond-ordered
phases coexist with a disordered phase as predicted by Domany, Schick and Swendsen\cite{DSS}. 
This last result generalizes the result of Shlosman and van Enter\cite{SvE}. 
We argue that these phenomena are quite general and should occur for a large
class of potentials.
\end{abstract}
\maketitle
\section{Introduction and main result}

A usual feature in  the Statistical Mechanics is the coexistence
of several phases at low temperature. Typically, the number of coexisting
pure phases decreases with increasing temperature up to some
critical temperature $\beta_c$ at which a single disordered phase
appears. This $q$-state Potts model for sufficiently large $q$ behaves 
slightly differently since it has  $q$ ordered phases below the transition 
temperature, but at  the transition it has $q+1$ phases, including a disordered 
phase \cite{KS,LMR}. Thus, the Potts model demonstrates that order and
disorder can coexist. 

There is a continued interest in exploring the variety of ordering phenomena 
occurring in models of classical continuous XY spins 
\cite{SDSS,BCK,NBCB}. In the present work, one of our motivations was to
look for the coexistence of ordered and disordered phases in such models (as
opposed to the previously studied Potts model, which is a discrete spin model).

An interesting situation of coexisting order and disorder 
was found to occur in a strongly non-linear $XY$ model \cite{DSS,BGH}.
For that model it was recently proved rigorously by Shlosman and van Enter
in Ref.~\onlinecite{SvE} that there is 
some temperature at which a first order phase transition occurs 
between a (ferromagnetic) bond ordered phase, which means that 
the nearest neighbors are close, and a bond  disordered phase, which means 
that the nearest neighbors are uncorrelated. This behavior resembles the 
Potts model with a magnetic field, which shows a unique ordered phase at low
temperature, and two coexisting phases at the transition. The main  difference is that
the $XY$ models has a continuous symmetry ($O(2)$), as opposed to Potts model where 
the symmetry is discrete. This continuous symmetry cannot be spontaneously broken 
by the Mermin-Wagner Theorem, which implies any long-range order is not
accompanied by a non-vanishing magnetization.

In this paper we introduce a new family of $O(2)$-models defined
on the two-dimensional square lattice, some of which exhibit a
second order phase transition at sufficiently low temperature that
coexists with a first order phase transition. Our family of models
depends on five parameters: three positive integers, $p, m$, and
$n$, and two non-negative coupling constants, $J$, and $K$:
\begin{equation}
H= 2J\sum_{[i,j]} \left[\cos{m\over 2} (\varphi_i
-\varphi_j)\right] ^{2p}+ K \sum_{[[i,k]]} \cos{n\over 2}
(\varphi_i -\varphi_k) \label{ham} \end{equation} 
where $\varphi_i
\in [0, 2\pi)$, and $[.,.]$ denotes nearest neighbors and
$[[.,.]]$ denotes diagonal neighbors. The model considered in Refs \onlinecite{DSS} and
\onlinecite{SvE}
is the special case $K=0$ and $m=1$. As we will only
consider $K >0$,  we can fix $K$ by rescaling the temperature;
we pick $K=1/4$. 

We prove that, for suitable values of the parameters, 
these models have both a first and a second order phase transition.
Our main result is the following theorem, in which we demonstrate the
occurrence of both phase transitions by means of estimates on 
on nearest neighbor and long range correlations.
The correlation functions with specified boundary
conditions (b.c.) are denoted by $\langle\cdot\rangle^{\rm b.c.}$.
We consider three types of boundary conditions: 1) $f$, for {\em ferromagnetic}, 
indicating parallel nearest neighbor spins, 2) $a$, for {\em antiferromagnetic},
which favors antiparallel nearest neighbor spins, and 3) $d$, for {\em disordered}, 
denoting the third phase, which appears at the transition temperature.
The precise meaning of the various boundary conditions is clarified later.

{\bf Theorem. }
{\em For sufficiently large $p$, $m=1$, and $n=2$, the models with
 Hamiltonian \eq{ham} have the following properties:

I) There exists an inverse temperature $\beta_t$, and functions
$\epsilon^{\rm b.c.} (\beta, p) <1$, and $\eta^{\rm b.c.} (\beta, p)<1$, such that:

a) for $\beta > \beta_t$, there is a first ordered phase
transition between the bond ferromagnetic phase and the bond
anti-ferromagnetic phase:
\begin{eqnarray}
\langle\cos (\varphi_i -\varphi_j)\rangle^f
&>& 1-\epsilon^f (\beta, p)\\
\langle\cos (\varphi_i -\varphi_j)\rangle^a
&<& -1+\epsilon^a(\beta, p)
\end{eqnarray}
for nearest neighbor pairs $[i,j]$.

b) at $\beta_t$ there is a first order phase transition between
the  bond ordered ferromagnetic  phase, the  bond
anti-ferromagnetic phase and the disordered phase:
\begin{eqnarray}
\langle\cos (\varphi_i -\varphi_j)\rangle^f
&>& 1-\eta^f(p)\\
\langle\cos (\varphi_i -\varphi_j)\rangle^a &<& -1+\eta^a (p)\\
\vert\langle\cos (\varphi_i -\varphi_j)\rangle^d \vert &<&
\eta^d (p)
\end{eqnarray}
II) At sufficiently low temperature there is  a second order phase
transition to phase with power law decay of the two point
correlation:
\begin{equation}
{C\over (1+ N) ^{{1\over 2\pi \beta'}}}\leq \langle\cos (\varphi_0
-\varphi_N)\rangle\leq { 1\over N^{{1\over p\beta\{2J+{1\over 2}
\}} }}
 \end{equation}
where $\beta' > {1\over 4\pi}$ and
$\beta'(\beta) \to \infty $ when $\beta \to \infty$, and $N$
denotes any site at distance $N$ from the origin.
}

We interprete our  result as follows:  two
phases coexist at low temperature, one    with a  local
ferromagnetic order, and one   with a local  anti-ferromagnetic
order,  but in large  distance spin waves     destroy both  the
long range ferromagnetic order  and the long range anti-ferromagnetic 
order leading to a second order  phase transition at some lower temperature.
The situation has to be opposed to the rotator model in two dimensions, 
which exhibits a second order phase transition at low
temperature, but   no first order phase transition \cite{BFL,MMP}.
Notice that  the Lebowitz inequalities \cite{Leb}  which  were used in Ref.~\onlinecite{BFL} to prove the uniqueness of the correlation functions in the
two dimensional rotator model are not valid in our model.

In the next section we describe the different phases of our model more precisely 
and introduce associated restricted ensembles used for the proof of the theorem.
The proof itself is given in a separate section. 

\section{The bond-ordered phases. Restricted ensembles}

We  want to prove that  below a temperature $\beta_t^{-1}$ two
two order phases coexist: one in which ferromagnetically ordered
bonds dominate and another in which antiferromagnetically
ordered bonds dominate. We also want to show that at the first-order
transition temperature, $\beta_t$,  three phases coexist: the two ordered 
phases plus a disordered phase. We will give the complete proof of 
this last statement, from which it will be then be clear how to prove the first.
First, we give some definitions and describe the ground states of the
model.

For the purpose of describing the ground states, a plaquette is called {\it  ferromagnetic} if
$ \varphi_i = \varphi_j $ for each nearest neighbor bond, $[i,j]$, of the plaquette,
and {\it anti-ferromagnetic } if $ \varphi_i - \varphi_j = \pi $ for each bond. 
We will extend these definitions to finite temperature a couple of paragraphs further down.

The  Hamiltonian  $H^{\{p,1,2\}}$ has  {\it two families of ground states}, 
each parametrized by an angle:

$\bullet$ the {\it ferromagnetic} ground states, in which all plaquettes are  
ferromagnetic;

$\bullet$  the {\it antiferromagnetic} ground states, which have only 
antiferromagnetic plaquettes.

It is not hard to see that these are the only ground states and it is easy to 
specify boundary conditions that select one of them.

Next, we define families of configurations that are close to one the ground
state configurations and that will carry most of the weight of the equilibrium states
for an appropriate range of temperatures. {\em Restricted ensembles} and the corresponding
partition functions can then be defined by summing over all configurations in each of these
families. For each bond $b=[i,j]$ 
three characteristic functions:
\begin{eqnarray}
\chi^f [\varphi_i -\varphi_j ] &=&
\left\{\begin{array}{rcl}1
&\mbox{if}& |\varphi_i -\varphi_j| < \epsilon\\
0&\mbox{else}&\end{array}\right.\\
 \chi^a [\varphi_i -\varphi_j ]
&=&\left\{\begin{array}{rcl}1
&\mbox{if}& |\varphi_i -\varphi_j-\pi| < \epsilon\\
0&\mbox{else}&\end{array}\right.\\
\chi^d [\varphi_i -\varphi_j ]& =& 1 - \chi^{\rm f}
[\varphi_i -\varphi_j ] - \chi^{\rm af} [\varphi_i
-\varphi_j ] 
\end{eqnarray}
Here again, and in the sequel, $f$ stands for ferromagentic, $a$ for antiferromagnetic,
and $d$ for disordered.

\begin{figure}[t]
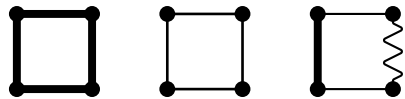

\pspicture(0,0)(3,3)
\psset{unit=.5cm,origin={4,0}}
\qdisk(1,1){3pt}\qdisk(1,3){3pt}\qdisk(3,1){3pt}\qdisk(3,3){3pt}
\psline[linewidth=3pt](1,1)(1,3)(3,3)(3,1)(1,1)
\psset{origin={0,0}}
\qdisk(1,1){3pt}\qdisk(1,3){3pt}\qdisk(3,1){3pt}\qdisk(3,3){3pt}
\psline[linewidth=1pt](1,1)(1,3)(3,3)(3,1)(1,1)
\psset{origin={-4,0}}
\qdisk(1,1){3pt}\qdisk(1,3){3pt}\qdisk(3,1){3pt}\qdisk(3,3){3pt}
\qline(1,1)(3,1)
\qline(1,3)(3,3)
\pszigzag[coilaspect=0,coilarm=.1,coilwidth=.3cm,linearc=.05]{-}(3,1)(3,3)
\psline[linewidth=3pt](1,1)(1,3)
\endpspicture
\caption{Three examples of  plaquettes. Thick lines denote ferromagnetic bonds (nearly
parallel spins), thin lines denote antiferromagnetic bonds (nearly antiparallel spins) and
a zigzag line is used to denote all other cases, which we call disordered bonds. 
The first two plaquettes are the ferro and
antiferromagnetic ground state plaquettes and will be labeled
$ffff$ and $aaaa$, respectively. The third plaquette is neither ferromagnetic,
antiferromagnetic, nor disordered and is labeled by the sequence $a d a\! f$. }\label{fig:plaquettes}
\end{figure}

The {\it specification of the bond is ferromagnetic} if $\chi^{f.}
[.]= 1$, {\it anti-ferromagnetic}  if $\chi^{\rm af} [.] = 1$, and
{\it disordered} if $ \chi^{\rm f} [.] = 1$. In the figures we will 
indicate a ferromagnetic bond by a thick line, an antiferromagnetic
bond by a thin line, and a disordered bond by a zigzag line.
The specification of the four bonds of a plaquette will be denoted
by $stuv$, where $s,t,u,v\in\{f,a,d\}$, and where the bonds are ordered
in clockwise fashion starting from the top horizontal bond.
As before, a ferromagnetic plaquette is one with specification
$ffff$, an antiferromagnetic plaquette has $aaaa$, etc.
(See Figure \ref{fig:plaquettes}). 

For a general configuration, we say that $\gamma$ is a {\it precontour} if it is is 
a maximal connected set of plaquettes that are neither ferromagnetic nor antiferromagnetic nor
disordered. A precontour  is composed  of  different  plaquettes
i.e.  plaquettes containing two or three different  bonds:  either
ferromagnetic, or anti-ferromagnetic or disordered.  A precontour
can be thick, and different  precontours can have the same
support. Notice that below $\beta_t^{-1}$ the precontours contain
typically  disordered plaquettes. 

The configurations of two plaquettes
are called {\it equivalent} if the specification of
each plaquette is the same.  An equivalent class of
configurations  in $V$ is a {\it restricted ensemble}.

\newsavebox{\afafP}
\savebox{\afafP}{\raisebox{-4pt}{\scaleboxto(.4cm,0){
\pspicture(0,0)(3,3)\psset{unit=1cm,origin={.7,0}}
\qdisk(1,1){6pt}\qdisk(1,3){6pt}\qdisk(3,1){6pt}\qdisk(3,3){6pt}
\psline[linewidth=1pt](1,3)(3,3)
\psline[linewidth=6pt](3,3)(3,1)
\psline[linewidth=1pt](3,1)(1,1)
\psline[linewidth=6pt](1,1)(1,3)
\endpspicture}}}
\newsavebox{\adaaP}
\savebox{\adaaP}{\raisebox{-4pt}{\scaleboxto(.4cm,0){
\pspicture(0,0)(3,3)\psset{unit=1cm,origin={.7,0}}
\qdisk(1,1){6pt}\qdisk(1,3){6pt}\qdisk(3,1){6pt}\qdisk(3,3){6pt}
\psline[linewidth=1pt](1,1)(3,1)
\psline[linewidth=1pt](1,3)(3,3)
\pszigzag[linewidth=1pt,coilaspect=0,coilarm=.1,coilwidth=.6cm,linearc=.05]{-}(3,1)(3,3)
\psline[linewidth=1pt](1,1)(1,3)
\endpspicture}}}
\newsavebox{\adadP}
\savebox{\adadP}{\raisebox{-4pt}{\scaleboxto(.4cm,0){
\pspicture(0,0)(3,3)\psset{unit=1cm,origin={.7,0}}
\qdisk(1,1){6pt}\qdisk(1,3){6pt}\qdisk(3,1){6pt}\qdisk(3,3){6pt}
\psline[linewidth=1pt](1,3)(3,3)
\pszigzag[linewidth=1pt,coilaspect=0,coilarm=.1,coilwidth=.6cm,linearc=.05]{-}(3,3)(3,1)
\psline[linewidth=1pt](3,1)(1,1)
\pszigzag[linewidth=1pt,coilaspect=0,coilarm=.1,coilwidth=.6cm,linearc=.05]{-}(1,1)(1,3)
\endpspicture}}}
\newsavebox{\aaddP}
\savebox{\aaddP}{\raisebox{-4pt}{\scaleboxto(.4cm,0){
\pspicture(0,0)(3,3)\psset{unit=1cm,origin={.7,0}}
\qdisk(1,1){6pt}\qdisk(1,3){6pt}\qdisk(3,1){6pt}\qdisk(3,3){6pt}
\psline[linewidth=1pt](1,3)(3,3)
\psline[linewidth=1pt](3,3)(3,1)
\pszigzag[linewidth=1pt,coilaspect=0,coilarm=.1,coilwidth=.6cm,linearc=.05](3,1)(1,1)
\pszigzag[linewidth=1pt,coilaspect=0,coilarm=.1,coilwidth=.6cm,linearc=.05](1,1)(1,3)
\endpspicture}}}
\newsavebox{\daddP}
\savebox{\daddP}{\raisebox{-4pt}{\scaleboxto(.4cm,0){
\pspicture(0,0)(3,3)\psset{unit=1cm,origin={.7,0}}
\qdisk(1,1){6pt}\qdisk(1,3){6pt}\qdisk(3,1){6pt}\qdisk(3,3){6pt}
\pszigzag[linewidth=1pt,coilaspect=0,coilarm=.1,coilwidth=.6cm,linearc=.05](1,3)(3,3)
\psline[linewidth=1pt](3,3)(3,1)
\pszigzag[linewidth=1pt,coilaspect=0,coilarm=.1,coilwidth=.6cm,linearc=.05](3,1)(1,1)
\pszigzag[linewidth=1pt,coilaspect=0,coilarm=.1,coilwidth=.6cm,linearc=.05](1,1)(1,3)
\endpspicture}}}

\begin{figure}[t]
\psset{unit=1cm,origin={0,0}}
\pspicture(0,0)(5,5)
\multips(1,1)(0,1){4}{\multips(0,0)(1,0){4}{\qdisk(0,0){2pt}}}
\multips(1,1)(0,1){4}{\qline(0,0)(3,0)}
\multips(1,1)(1,0){4}{\psline[linewidth=3pt](0,0)(0,3)}
\endpspicture
\caption{The universal contour $U(a\!f\!a\!f)$ obtained from
the plaquette specification \usebox{\afafP} by repeated reflections.}\label{fig:afaf}
\end{figure}

\begin{figure}[t]
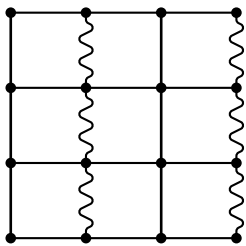

\psset{unit=1cm,origin={0,0}}
\pspicture(0,0)(5,5)
\multips(1,1)(0,1){4}{\multips(0,0)(1,0){4}{\qdisk(0,0){2pt}}}
\multips(1,1)(0,1){4}{\qline(0,0)(3,0)}
\multips(1,1)(2,0){2}{\psline[linewidth=1pt](0,0)(0,3)}
\multips(2,1)(2,0){2}{\multips(0,0)(0,1){3}{
\pszigzag[coilaspect=0,coilarm=.1,coilwidth=.3cm,linearc=.05]{-}(0,0)(0,1)}}
\endpspicture
\caption{The universal contour $U(adaa)$ obtained from
the plaquette specification \usebox{\adaaP} by repeated reflections.}\label{fig:adaa}
\end{figure}

We  define three  kinds of restricted ensembles:  1) {\it the
restricted  ferromagnetic  ensemble $R^f_V$} built
with equivalent ferromagnetic plaquettes in $V$; 2)
{\it the    restricted  antiferromagnetic  ensemble
$R^a_V$} built with equivalent
antiferromagnetic plaquettes; 3)
 {\it the restricted disordered  ensemble $R^d_V$} built
with equivalent disordered  plaquettes. 

There is a one to one correspondence between the restricted ensembles $R^f_V$   and
$R^a_V$. To see this, consider the decomposition of the lattice into the two 
standard sublattices A and B, where the nearest neighbors of each site in B (A) 
are in A (B). The, with each ferromagnetic configuration there corresponds an antiferromagnetic one obtained by adding $\pi$  to each spin on the sites of A.

The  {\it  restricted  partition functions}   $\Xi_V(f)$,  $\Xi_V(a)$, and $\Xi_V(d)$, are the
partition functions with the configuration sum restricted to $R^f_V$, $R^a_V$,
and $R^d_V$, respectively.

The one-to-one correspondence between $R^f_V$   and
$R^a_V$, implies that $\Xi_V^{\rm f} =\Xi _V^{\rm af}$.  


 A {\it contour} $\Gamma$ is the set of configurations,
 which are equivalent to a precontour's configuration $\gamma$. A
 contour $\Gamma$ is specified by:
1) a geometric support  $S(\Gamma)$,  which is  a connected set of
plaquettes;
2)   a family of equivalent configurations  defined on
$S(\Gamma)$.

\section{The proof}

{\bf  Proof of I.a and I.b}. The  existence of a first order phase
transition \cite{KS} is  a consequence of   the three conditions:

a) both the probability of an ordered plaquette of an
antiferromagnetic plaquette are small at high temperature,

b) the probability of a disordered plaquette is small at low
temperature,  c) the probability of the  contours are small at
every  temperature.    The  proof of   I.a) and I.b) will be
achieved  via chess-board estimates, which require that the
Hamiltonian is {\it reflection positive} \cite{FSS,FILSa,FILSb}.  We  expand
the Hamiltonian.
\begin{eqnarray}
H&=& J \sum_{[i,j], m=1,2,\ldots q)} A_m 
\cos 2m (\varphi_i -\varphi_j)\nonumber \\
&&\quad+{1\over 4} \sum_{[[i,k]] \in V} \cos(\varphi_i
-\varphi_k)\label{hamRP}\end{eqnarray}

Here, the $A_m$  are positive integers.  Obviously the Hamiltonian \eq{hamRP} is
reflection positive with respect to the axis of the lattice. For simplicity we choose $J=1$.

We prove directly part I.b), because I.a)   follows by
symmetry. Peierls  bounds via chessboard
estimates are obtained using  {\it universal contours},
which are obtained from  a  plaquette configuration $stuv$, by
reflections through the axes of the lattice. See Figures 2 and 3
for examples of universal configurations $U(stuv)$, which
should be thought of as equivalence classes of configurations.

The probability of a plaquette configuration $stuv$ is given via the following  formula, 
which  follows from successive applications  of the Schwartz inequality \cite{FILSa,FILSb}. 
$$
\langle stuv\rangle>_V \leq \left[ {\Xi_V [U(stuv)]\over \Xi_V}\right]^{{1\over
|V|}} \eqno (7)$$ First we need to prove a lower  bound  for
$\Xi_V$, which is similar to the  one used in Ref.~\onlinecite{SvE}.
\begin{equation}
 \Xi_V \geq   \left[ {1\over C \sqrt p}\right]^{|V|}
\exp\beta |V| [ {5\over 2} -O({1\over C^2})] + \left[ 1-{4C \over
\sqrt p}\right]^{{|V|\over 2}}  \label{LB1} 
\end{equation}
We briefly recall the argument leading to \eq{LB1}.

\begin{figure}[t]
\psset{unit=1cm,origin={0,0}}
\pspicture(0,0)(5,5)
\multips(1,1)(0,1){4}{\multips(0,0)(1,0){4}{\qdisk(0,0){2pt}}}
\multips(1,1)(0,1){4}{\qline(0,0)(3,0)}
\multips(1,1)(1,0){4}{\multips(0,0)(0,1){3}{
\pszigzag[coilaspect=0,coilarm=.1,coilwidth=.3cm,linearc=.05]{-}(0,0)(0,1)}}
\endpspicture
\caption{The universal contour $U(adad)$ obtained from
the plaquette specification \usebox{\adadP} by repeated reflections.}\label{fig:adad}
\end{figure}

\begin{figure}[t]
\psset{unit=1cm,origin={0,0}}
\pspicture(0,0)(5,5)
\multips(1,1)(0,1){4}{\multips(0,0)(1,0){4}{\qdisk(0,0){2pt}}}
\multips(1,2)(0,2){2}{\qline(0,0)(3,0)}
\multips(2,1)(2,0){2}{\qline(0,0)(0,3)}
\multips(1,1)(2,0){2}{\multips(0,0)(0,1){3}{
\pszigzag[coilaspect=0,coilarm=.1,coilwidth=.3cm,linearc=.05]{-}(0,0)(0,1)}}
\multips(1,1)(0,2){2}{\multips(0,0)(1,0){3}{
\pszigzag[coilaspect=0,coilarm=.1,coilwidth=.3cm,linearc=.05]{-}(0,0)(1,0)}}
\endpspicture
\caption{The universal contour $U(aadd)$ obtained from
the plaquette specification \usebox{\aaddP} by repeated reflections.}\label{fig:aadd}
\end{figure}

\begin{figure}[t]
\psset{unit=1cm,origin={0,0}}
\pspicture(0,0)(5,5)
\multips(1,1)(0,1){4}{\multips(0,0)(1,0){4}{\qdisk(0,0){2pt}}}
\multips(2,1)(2,0){2}{\qline(0,0)(0,3)}
\multips(1,1)(2,0){2}{\multips(0,0)(0,1){3}{
\pszigzag[coilaspect=0,coilarm=.1,coilwidth=.3cm,linearc=.05]{-}(0,0)(0,1)}}
\multips(1,1)(0,1){4}{\multips(0,0)(1,0){3}{
\pszigzag[coilaspect=0,coilarm=.1,coilwidth=.3cm,linearc=.05]{-}(0,0)(1,0)}}
\endpspicture
\caption{The universal contour $U(dadd)$ obtained from
the plaquette specification \usebox{\daddP} by repeated reflections.}\label{fig:dadd}
\end{figure}

The first term is obtained  by integrating over all the
configurations  such that  $|\varphi_i | \leq {1\over C \sqrt{p}} $
for all $i\in V$, the second one by integrating over all the
configurations over one N\'eel sublattice  which satisfy  $|
\varphi_i -\varphi_j| > {1\over C \sqrt p}$ for all nearest
neighbors.

Secondly  we  derive an upper bound for the quantities $
\Xi_V [U(stuv)]$  for the different restricted ensembles $P(D)$. The
proofs   of the conditions a) and b) are straightforward, so we
focus on the proof of c), which is to prove  that the
probabilities of  the various plaquettes occurring in contours are  small at
every temperature. Let  $\beta_0$ be  defined  by: 
\begin{equation}
 e^{\beta_0}=
\left[ pC^2(1-{4C\over \sqrt{p}})\right]^{[{1\over 5}+O[{1\over
C^2}] }   \label{9}
\end{equation}

We  have to estimate the probability of the contours with
different specifications.

$\bullet$  {\sl Two opposite bonds with $\chi^f =1$ and two
opposite bonds with $\chi^a =1$}, i.e., a plaquette with specification $a\!f\!a\!f$.
The universal contour $U(a\!f\!a\!f)$  is built from  ferromagnetic vertical and
horizontal bonds with even coordinates, and from
anti-ferromagnetic vertical and  horizontal bonds with odd
coordinates as shown in Figure \ref{fig:afaf}. 
We get the upper bound for $ \Xi_V [U(a\!f\!a\!f)]$  by
noticing that the energies of the diagonal   potentials are zero:
$$
\Xi_V [U(a\!f\!a\!f)] \leq \left[ {1\over C \sqrt{p}}\right]^{|V|} \exp[\beta |V|
\left[2 +  O({1\over C^2})\right]
$$

a) We get for $\beta > \beta_0$:

\begin{eqnarray}
\lefteqn{\langle\usebox{\afafP}\rangle}\nonumber\\
 & =&\left[ { \Xi_V [U(a\!f\!a\!f)]\over
\Xi_V}
\right]^{{1\over |V|}} \nonumber\\
&\leq& \left[ {\left[  {1\over C \surd p}\right]^{ |V|}  \exp\beta
|V| [  2 +    O({1\over C^2}) ] \over \left[ {1\over C \sqrt{p}}\right]^{|V|} \times \exp\beta |V| ( {5\over 2}-O({1\over C^2}) )
+ \left[ 1-{4C\over \sqrt{p}}\right]^{{|V|\over 2}}} \right]^{{2\over |V|}}\nonumber\\
& \leq& \exp-\beta\left\{ {1 \over 2}   -O({1\over C^2})\right\}\nonumber\\
&\leq& \left[ {1\over C \surd p}\right]^{[{1\over 5}-O({1\over
C^2})]}\times  \left[1+{4C\over \sqrt{p}}\right]^{{1\over
10}}\label{eq:11}
\end{eqnarray}

b) For $\beta < \beta_0$ we obtain:
\begin{eqnarray*}
\langle\usebox{\afafP}\rangle  &\leq& {1\over C \sqrt{p}} {\left[
pC^2(1-{4C\over \sqrt{p}})\right]^{[{2\over 5}+O({1\over C^2})]}
\over \left[1-{4C\over \sqrt{p}}\right]^{{1\over 2}}}\\
 &\leq&
\left[{1\over C \surd p}\right]^{({1\over 5}-O[{1\over
C^2}])}\times \left[1+{4C\over \sqrt{p}}\right]^{{1\over 10}}
\end{eqnarray*}
In both cases the expectations can be made small for $p$ large
enough compared to $C$.   It is obvious that the upper bounds for the
expressions of $\langle\usebox{\afafP}\rangle$ computed at  low and at high
temperature have the same dependence on $C$
and $p$. This holds in general and therefore we will compute only one case for the other
specifications.

$\bullet$ {\sl One disordered bond and three ordered bonds.} The nature of the ordered bonds, $f$ or $a$, produces only minor changes. Therefore, let us focus on one instances, say the universal
contour $U(adaa)$ which is constructed from ordered horizontal bonds and  even
vertical bonds,  and from disordered vertical odd bonds, as shown in Figure
\ref{fig:adaa}. For the upper bound for $ \Xi_V [U(adaa)]$, we notice that the energy
of  the odd vertical  bonds is zero:
$$
\Xi_V [U(adaa)] \leq \left[{1\over C \sqrt{p}}\right]^{|V|} \exp\beta
|V|  [  2 +  O({1\over C^2})]\, ,
$$
which leads to the upper bound: 
\begin{equation}
\langle\usebox{\adaaP}\rangle  \leq   \left[{1\over C \sqrt{p}}\right]^{({1\over 5}-O[{1\over C^2}])}\times \left[1+{4C\over\sqrt{p}}\right]^{{1\over 5}} 
\label{eq:12}\end{equation}

$\bullet$  {\sl Two opposite  disordered bond and two opposite
ordered bonds.} The universal contour $U(adad)$ is built from
ordered horizontal bonds,  and from disordered vertical bonds
as shown in Figure \ref{fig:adad}. We
get the  upper bound for the $ \Xi_V [U(adad)]$ by noticing that
the energy of  the  vertical bonds is zero:
\begin{equation}
\Xi_V [U(adad)] \leq \left[{1\over C \sqrt{p}}\right]^{|V|} \exp
\beta |V|  ( {3\over 2} +  O({1\over C^2})  \label{eq:13}
\end{equation}

We get the upper bound:
\begin{equation}
\langle\usebox{\adadP}\rangle  \leq   \left[ {1\over C \surd p}\right]^{({2\over 5}-O[{1\over C^2}])}\times
\left[1+{4C\over \sqrt{p}}\right]^{{1\over 5}} 
\label{eq:14}\end{equation}

$\bullet$ {\sl Two adjacent  disordered bonds and two adjacent
ordered bonds.} The universal contour $U(aadd)$ is built from
ordered horizontal even bonds  and even vertical bonds, and from
disordered  vertical odd and the horizontal  bonds as shown in Figure
\ref{fig:aadd}. We get the
upper bound for $ \Xi_V [U(aadd)]$ the restricted partition
function by noticing that the energy of  the odd vertical and
horizontal  bonds are zero, and that there   are entropy
contributions from the vertices with odd coordinates:
\begin{equation}
 \Xi_V [U(aadd)] \leq \left[{1\over C \sqrt{p}}\right]^{{3\over 4}|V|}
 \exp[\beta |V|  \left[{3\over 2} +  O({1\over C^2})\right] \label{eq:15}
 \end{equation}

We get the upper bound:
\begin{equation}
\langle\usebox{\aaddP}\rangle  \leq    \left[ {1\over C \sqrt{p}}\right]^{({3\over 20}-O[{1\over C^2}])}\times
\left[1+{4C\over \sqrt{p}}\right]^{{1\over 5}} \label{eq:16}
\end{equation}

$\bullet$ {\sl Three disordered bond and one  ordered bonds.} The
universal contour $U(dadd)$ is built from disordered horizontal
bonds  and even vertical bonds, and  from ordered vertical odd
bonds as shown in Figure \ref{fig:dadd}. 
We get the  upper bound for $ \Xi_V [U(dadd)]$ by noticing
that the energy of  the horizontal  and the odd vertical and bonds
is zero, and that there are entropy contributions  from the
vertices with even  ordinates:
\begin{equation}
\Xi_V [U(dadd)] \leq \left[{1\over C \sqrt{p}}\right]^{{1\over
2}|V|} \exp\beta |V|  [1 +  O({1\over C^2})] \label{eq:17}
\end{equation}

We get the upper bound: 
\begin{equation}
\langle\usebox{\daddP}\rangle \leq   \left[{1\over C
\sqrt{p}}\right]^{({1\over 10}-O[{1\over C^2}])}\times
\left[1+{4C\over \sqrt{p}}\right]^{{3\over 10}} \label{eq:18}
\end{equation}

For $p\gg C\gg 1$, the probability of each specification of a contour's
plaquette can be made small enough.  Finally  we  prove the
Peierls condition: first we sum over all the specifications of the
contours    with a fixed support using the estimates \eq{eq:11}, \eq{eq:12},
\eq{eq:14}, \eq{eq:16}, and \eq{eq:18}; next  we use the Koenigsberg's lemma to get an
upper bound on  the entropy for the  contours. This concludes our proof
of part I of the Theorem.

{\bf Proof  of II)} The upper bound on the two points
correlation function is proved via a spin wave argument, the lower
bound is deduced  from  the deep result of Fr\"ohlich and Spencer
for the two dimensional classical XY model \cite{FS}. We now provide
the details. 

{\bf II.a) The lower bound}

We  use  the  Ginibre inequalities \cite{Gin}, which are valid for  ferromagnetic Hamiltonians.
\begin{eqnarray}
\lefteqn{\langle \cos (m_1[\varphi_i -\varphi_j])\times  \cos
(m_2[\varphi_k -\varphi_l])\rangle^{H^{p}_V}}\nonumber \\
&\geq& \langle  \cos (m_1[\varphi_i -\varphi_j])\rangle^{H^{p}_V}
\langle \cos (m_2[\varphi_k -\varphi_l])\rangle^{H^{p}_V}\label{eq:gin}
\end{eqnarray}
We first start by removing all the terms contained in the Hamiltonian with
the exception of the  diagonal interactions, then we are left with
two independent rotator models, one on each of the sublattices.

First, suppose $i$ and $j$ are on the same sublattice. Then we 
use the remaining interactions on that sublattice. We are left with 
the Hamiltonian:  $H_V^{R} =:
\sum_{\{[[i,k]] \in V\}} \cos ([\varphi_k -\varphi_i])$, and we have
$$
\langle \cos (\varphi_0 -\varphi_N)\rangle^{H^{p}_V}
\geq  \langle\cos (\varphi_0 -\varphi_N)\rangle^{H^R_V}
$$
Next we apply the result of Fr\"ohlich and Spencer \cite{FS}, Theorem C,  for the two
dimensional rotator model to get a lower bound.
\begin{equation}
\langle\cos (\varphi_0 -\varphi_N)\rangle^{H^{p}_V} \geq   {C\over (1+ N)
^{{1\over 2\pi \beta'}}}  \label{eq:21}
\end{equation}
where $\beta' > {1\over 4\pi}$ and $\beta'(\beta) \to \infty $ when $\beta \to \infty$.

If $i$ and $j$ belong to different sublattices, the same lower bound of \eq{eq:21}
holds with a slightly smaller constant $C$, again by using the Ginibre inequality
\eq{eq:gin} with $j=l$ and $k$ and $j$ replaced by a nearest neighbor of $j$.

\begin{figure}[t]
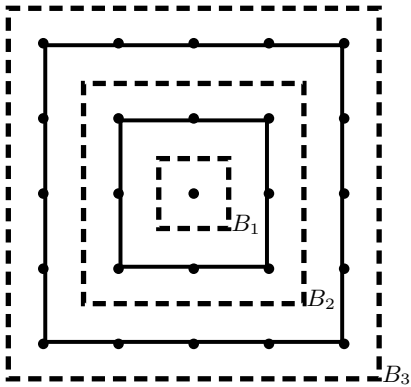

\psset{unit=1cm,origin={0,0}}
\pspicture(0,0)(7,7)
\multips(1,1)(0,1){5}{\multips(0,0)(1,0){5}{\qdisk(0,0){2pt}}}
\psframe[linestyle=dashed,linewidth=2pt](0.5,0.5)(5.5,5.5)
\psframe[linestyle=dashed,linewidth=2pt](1.5,1.5)(4.5,4.5)
\psframe[linestyle=dashed,linewidth=2pt](2.5,2.5)(3.5,3.5)
\psframe[linestyle=solid,linewidth=1.5pt](1,1)(5,5)
\psframe[linestyle=solid,linewidth=1.5pt](2,2)(4,4)
\put(3.5,2.5){$B_1$}
\put(4.5,1.5){$B_2$}
\put(5.5,0.5){$B_3$}
\endpspicture
\caption{The nested square boxes $B_l$ used in the argument for the upper bound
II.b. are shown in dashed lines. The solid bonds are the new strongly
ferromagnetic interactions introduced in the proof.}\label{fig:onion}
\end{figure}

{\bf II.b)  The upper bound.}
The upper bound  for the two points correlation function is 
obtained  by  using a well known spin wave
argument. We use a nice argument that we learned from H. Kunz\cite{Kun}. 
It starts by defining a sequence of squares (or hypercubes in higher dimensions) 
and label them $B_1,B_2, \ldots$, as shown in Figure \ref{fig:onion}. Each
$B_l$ is centered at the origin and its boundaries can be taken to intersect nearest 
neighbor bonds.
We define the shells $C_l = B_{l+1}\setminus B_l$, each shell
$C_l$ contains a square $R_l$ of the lattice, the nearest
vertices contained in $R_l$ are labelled by a couple $[l, i(l)]$,
where $ i(l) $ denotes the coordinates along $R_l$. We want to
identify the continuous spins   contained in each shell $R_l$  by
using  Ginibre's  inequality \eq{eq:gin}.  To do so, we  add to the
Hamiltonian $H_V$ the ferromagnetic interactions:
$\sum_{l=1}^{l=N} \sum_{i(l)} J_l \cos (\varphi_{ i+1(l)}
-\varphi_{i(l)})$, where the $J_l$ is are positive. The
correlations are increasing by Ginibre's inequality. Next  we let
$J_l = \infty$, the correlations are again increasing:
\begin{eqnarray*}
\lefteqn{\langle \cos (\varphi_0 -\varphi_N)\rangle_V^{H^p}} \\
 &\leq&
{1\over Z_L^{cr}} \times \int_0^{2\pi} d\varphi_1\cdots
\int_0^{2\pi} d\varphi_{(N+1)}\prod_{j=1}^{j=N}\times\\
&& \exp\big[ 4jJ \sum_{m=1,2,\ldots q} A_m 
\cos 2m(\varphi_j -\varphi_{j+1})\\
&& + {1\over 2} \cos (\varphi_j -\varphi_{j+1} )\big] \times
\exp\{i(\varphi_j -\varphi_{j+1})
\end{eqnarray*}

Where $ Z_L^{cr}$ is the normalization factor. The LHS is the
two points correlation function  corresponding to  a one
dimensional model, each term of the product can be computed, we
get the following upper bound:
$$
\langle \cos (\varphi_0 -\varphi_N)\rangle^{H^{p}} \leq N^{-1/(2p\beta\{2J+{1\over 2}\})}
$$

\section{Conclusion}

The coexistence  between  second order and first order phase
transitions at the same temperature  should be a fairly general
phenomena  in two-dimensional models with a continuous symmetry. 
For example it   occurs  in several
non linear  two dimensional XY models defined by (1) for other
values of the pair  $m,n$. In a more general setting, we have  to
consider  potentials with continuous symmetry, which are   peaked
and  with   large flat parts to produce  entropy. The problem will
be to characterize the criticality, which in our case uses \cite{FS},  a
generalization of the  work of Aizenman could be an alternative 
\cite{Aiz}. We expect that  the same situation should hold  for some
quantum models.

\begin{acknowledgments} A.M. would like to thank S. Shlosman for
helpful discussions, and UC Davis for an invitation during which a
part of this paper was done. 
B.N. acknowledges interesting conversations with L. Chayes, M. Biskup, and R.P. Singh. 
This material is based upon work supported by the National Science Foundation under
Grant No. DMS0303316.
\end{acknowledgments}

\end{document}